\documentclass[entropy,journal,article,accept,oneauthor,pdftex,10pt,a4paper]{mdpi} 
\firstpage{1} 
\pubvolume{xx}
\issuenum{1}
\articlenumber{1}
\pubyear{2018}
\copyrightyear{2018}
\history{}

\newcommand{\E}{\mathbb{E}}
\newcommand{\R}{\mathbb{R}}
\renewcommand{\d}{\,\mathrm{d}}
\let\phi\varphi
\let\tilde\widetilde

\allowdisplaybreaks

\Title{R\'enyi Entropy Power Inequalities\\\emph{via} Normal Transport and Rotation}

\Author{Olivier Rioul\orcidA{}} 

\AuthorNames{Firstname Lastname, Firstname Lastname and Firstname Lastname}

\address{%
LTCI, Télécom ParisTech, Université Paris-Saclay, 75013, Paris, France; olivier.rioul@telecom-paristech.fr\\
\'Ecole Polytechnique, Université Paris-Saclay, 91128 Palaiseau, France; olivier.rioul@polytechnique.edu}

\corres{Correspondence: e-mail@e-mail.com; Tel.: +x-xxx-xxx-xxxx}

\abstract{%
Following a recent proof of Shannon's entropy power inequality (EPI), a comprehensive framework for deriving various EPIs for the Rényi entropy is presented that uses transport arguments from normal densities and a change of variable by rotation. Simple arguments are given to recover the previously known Rényi EPIs and derive new ones, by unifying a multiplicative form with constant~$c$ and a modification with exponent $\alpha$ of previous works. 
In particular, for log-concave densities, we obtain a simple transportation proof of a sharp varentropy bound.
}


\usepackage{booktabs}

\begin{document}




\section{Introduction}

The entropy power inequality (EPI) dates back to Shannon's seminal paper~\cite{Shannon48} and has a long history~\cite{Rioul11}. The link with the Rényi entropy was first made by Dembo, Cover and Thomas~\cite{DemboCoverThomas91} in connection with Young's convolutional inequality with sharp constants,
where Shannon's EPI is obtained by letting the Rényi entropy orders tend to one~\cite[Theorem~17.8.3]{CoverThomas06}.

The Rényi entropy~\cite{Renyi61} was first defined as a generalization of Shannon's entropy for \emph{discrete} variables, when looking for the most general definition of information measures that would preserve the additivity for independent events. It has found many applications such as source coding~\cite{Campbell65}, hypothesis testing~\cite{BenBassatRaviv78}, channel coding~\cite{Arimoto75} and guessing~\cite{Arikan96}. The (differential) Rényi entropy considered in this paper (Definition~\ref{RenyiEntropy} below) generalizes the (differential) Shannon's entropy for continuous variables. It was first considered in~\cite{DemboCoverThomas91} to make the transition between the entropy power and the Brunn-Minkowski inequalities. It has also been applied to deconvolution problems~\cite{Erdogmusetal04}. A definition of the Renyi entropy-power itself appears in~\cite{SavareToscani14}, that is essentially Definition~\ref{RenyiEntropyPower} below.

Recently there has been significant interest in R\'enyi entropy power inequalities for several independent variables\footnote{The recent survey~\cite{MadimanMelbourneXu17} is recommended to the reader for various recent developments on forward and reverse entropy power inequalities.}.
Bobkov and Chistyakov~\cite{BobkovChistyakov15} extended the classical Shannon's EPI to the Rényi entropy by incorporating a multiplicative constant that depends on the order of the Rényi entropy. Ram and Sason~\cite{RamSason16} improved the value of the constant by making it depend also on the number of variables. Even more recently, Bobkov and Marsiglietti~\cite{BobkovMarsiglietti17} proved another modification of the EPI for the Rényi entropy for two independent variables, with a power exponent parameter $\alpha$ whose value was further improved by Li~\cite{Li18}. All these EPIs were found for Rényi entropies of orders $>1$. The $\alpha$-modification of the Rényi EPI was extended to orders $<1$ for two independent variables having log-concave densities by Marsiglietti and Melbourne~\cite{MarsigliettiMelbourne18}. The starting point of all the above works was Young's strengthened convolutional inequality. 

Also recently, Shannon's original EPI was given a simple proof~\cite{Rioul17} using a simple transport argument from normal variables and a change of variable by rotation. 
In this paper, we exploit these ingredients, described in the following lemmas, to establish all the above mentioned  Rényi EPIs and derive new ones.

\begin{Notation}
Throughout this article the considered $n$-dimensional zero-mean random variables $X\in\R^n$ admit a density which is implicitly assumed continuous inside its support. We write $X\sim f$ if $X$ has density $f$ and write $X^*\sim \mathcal{N}(0,\mathbf{K})$ if $X^*$ is normally distributed with $n\times n$ covariance matrix $\mathbf{K}$.
\end{Notation}
\begin{Lemma}[Normal Transport]\label{gt}
Let $X^*\sim \mathcal{N}(0,\sigma^2\mathbf{I})$. There exists a diffeomorphism $T:\R^n\to\R^n$ such that $X=T(X^*) \sim f$.
Moreover, $T$ can be chosen such that its Jacobian matrix $T'$ is (lower) triangular with positive diagonal elements.
\end{Lemma}
\noindent This lemma is known in optimal transport theory as an application of the Knothe-Rosenblatt map~\cite{Rosenblatt52,Knothe57}. Two different proofs are given in~\cite{Rioul17}. The proof is very simple for one-dimensional variables~\cite{Rioul17a}, where $T$ is just an increasing function with continuous derivative $T'>0$.

\begin{Lemma}[Normal Rotation~\cite{Rioul17}]\label{gr}
If $X^*,Y^*$ are i.i.d.\footnote{Independent and identically distributed.} normal, then for any $0<\lambda<1$, the rotation%
\begin{equation}
\begin{cases}
\tilde{X} \;=\;\hphantom{\sqrt{1-}}\sqrt{\lambda}\;X^*  &\!\!\!\!+ \sqrt{1-\lambda}\;Y^*
\\
\tilde{Y}\;=\;-\sqrt{1-\lambda}\;X^*  &\!\!\!\!+ \hphantom{\sqrt{1}}\sqrt{\lambda}\;Y^*
\end{cases} 
\end{equation}
yields i.i.d.\@ normal variables $\tilde{X},\tilde{Y}$. 
\end{Lemma}
\noindent Notice that the starred variables can be expressed in terms of the tilde variables by the inverse rotation
\begin{equation}\label{invrot}
\begin{cases}
X^* \;=\;\hphantom{\sqrt{-}}\sqrt{\lambda}\;\tilde{X}  &\!\!\!\!- \sqrt{1-\lambda}\;\tilde{Y}
\\
Y^*\;=\;\sqrt{1-\lambda}\;\tilde{X}  &\!\!\!\!+ \hphantom{\sqrt{1}}\sqrt{\lambda}\;\tilde{Y}.
\end{cases} 
\end{equation}
The proof of Lemma~\ref{gr} is trivial considering covariance matrices. A deeper result states that this property of remaining i.i.d.\@ by rotation characterizes the normal distribution---this is known as Bernstein's lemma (see e.g.,~\cite[Chap.~5]{Bryc95},~\cite[Lemma~4]{Rioul17a}). This explains why one obtains equality in the EPI only for normal variables (see~\cite{Rioul17a} for more details).

This article is a revised, full version of what was presented in part in a previous conference communication~\cite{Rioul17b}. It is organized as follows. Preliminary definitions and known properties are presented in Section~\ref{secprelim}.  Section~\ref{iisec} derives a crucial ``information inequality'' for Rényi entropies that enjoys a transformational invariance. The central result is in Section~\ref{repi1sec}, where the first version of the Rényi EPI by Dembo, Cover and Thomas is proved using the ingredients of Lemmas~\ref{gt} and~\ref{gr}. All previously known Rényi EPIs for finite orders---and new ones---are then derived using a simple method in Section~\ref{repi2sec}. Section~\ref{conclusion} concludes.

\section{Preliminary Definitions and Properties}\label{secprelim}

Throughout this article we consider exponents $p>0$ with $p\ne 1$. 
The following definition is well known and used e.g., in H\"older's inequality.

\begin{Definition}[Conjugate Exponent]
The \emph{conjugate exponent} of $p$ is
\begin{equation}
p'=\frac{p}{p-1}, 
\end{equation}
that is, the number $p'$ such that  
\begin{equation}
\frac{1}{p}+\frac{1}{p'}=1.
\end{equation}
\end{Definition}
\begin{Remark}
There are two situations depending on whether $p'$ is positive or negative, as summarized in the following table.
\begin{center}
\begin{tabular}{|c|c|c|}
\toprule
$p>1$ &\raisebox{-1ex}{or}& $0<p<1$ \\
$p'>1$ && $p'<0$ \\
\bottomrule 
\end{tabular} 
\end{center} 
\end{Remark}

\begin{Definition}[Rényi Entropy]\label{RenyiEntropy}
If $X$ has density $f\in L^p(\R^n)$, its R\'enyi entropy of order $p$ is defined by
\begin{align}
h_p(X)&=\frac{1}{1-p}\log \int_{\R^n} f^p(x) \d x \\
&= -p' \log\|f\|_p \label{hpdef2}
\end{align}
where $\|f\|_p$ denotes the $L^p$ norm of $f$.
\end{Definition}

It is known that the limit as $p\to 1$ is the Shannon entropy
\begin{equation}
h_1(X)=  -\int_{\R^n} f(x) \log f(x) \d x.
\end{equation}
The Rényi entropy enjoys well-known properties similar to those of the Shannon entropy, which are recalled here for completeness.
\begin{Lemma}[Scaling Property]\label{scaling}
For any $a\ne0$,
\begin{equation}
h_p(aX)=h_p(X) + n\log|a|.
\end{equation}
\end{Lemma}

\begin{proof}
Making a change of variables, $h_p(aX)=\frac{1}{1-p}\log\int \bigl(\frac{1}{|a|^n}f(\frac{x}{a}) \bigr)^p \d x =\frac{1}{1-p}\log\int f^p(\frac{x}{a}) \frac{\d x}{|a|^n} + \frac{1}{1-p}\log|a|^{n(1-p)}=h_p(X)+n\log|a|$.
\end{proof}
One recovers the usual scaling property for the Shannon entropy by letting $p\to 1$.

\begin{Lemma}[Rényi Entropy of the Normal] \label{gausshp} 
If $X^*\sim \mathcal{N}(0,\mathbf{K})$ for some nonsingular covariance matrix $\mathbf{K}$, then
\begin{equation}
h_p(X^*) = \frac{1}{2} \log ((2\pi)^n|\mathbf{K}|) + \frac{n}{2} p' \frac{\log p}{p}
\end{equation}
where $|\cdot|$ denotes the determinant. In particular for $X^*\sim \mathcal{N}(0,\sigma^2\mathbf{I})$,
\begin{equation}
h_p(X^*) = \frac{n}{2} \log (2\pi\sigma^2) + \frac{n}{2} p' \frac{\log p}{p}.
\end{equation}
\end{Lemma}
\begin{proof}
By direct calculation,  
$
h_p(X^*)=\frac{1}{1-p}\log {\displaystyle\int} \Bigl( \frac{\exp (-\frac{1}{2}x^t\mathbf{K}^{-1}x)}{\sqrt{(2\pi)^n|\mathbf{K}|}} \Bigr)^p \d x
=\frac{1}{1-p}\log \frac{ \sqrt{(2\pi)^n|\mathbf{K}|p^{-n}}  }{ \sqrt{(2\pi)^n|\mathbf{K}|}^p  }
= \frac{1}{2} \log ((2\pi)^n|\mathbf{K}|) - \frac{n}{2} \frac{\log p}{1-p} $.
\end{proof}
Again one recovers the Shannon entropy of a normal variable by letting $p\to 1$ (then $p'\frac{\log p}{p}\to \log e$). 

The following notion of escort distribution~\cite{Bercher09,Bercher12} is useful in the sequel.
\begin{Definition}[Escort Density~{\cite[\S~2.2]{Bercher12}}]\label{escort}
If $f\in L^p(\R^n)$, its \emph{escort density} of exponent $p$  is the density defined by
\begin{equation}
f_p(x) = \frac{f^p(x)}{\int_{\R^n} f^p(x) \d x}.
\end{equation}
In other words $f_p={f^p}/{\|f\|^p_p}$ where $\|f\|_p$ denotes the $L^p$ norm of $f$.
We also use the notation $X_p$ to denote the corresponding escort random variable with density $f_p$.
\end{Definition}

\begin{Lemma}[Monotonicity Property]\label{mono}
If $p<q$ then  $h_p(X)\geq h_q(X)$ with equality if and only if $X$ is uniformly distributed.
\end{Lemma}

\begin{proof}
Let $p\ne 1$ and assume that $f\in L^q(\R^n)$ for all $q$ in a neighborhood of $p$ so that one can freely differentiate under the integral sign:
\begin{align}
\frac{\partial}{\partial p} h_p(X) &=  \frac{1}{(1-p)^2}\log \int f^p + \frac{1}{1-p} \frac{\int f^p \log f }{ \int f^p}\\
&=\frac{1}{(1-p)^2} \bigl( \log \int f^p + \int f_p \log f^{1-p}\bigr)\\
&=\frac{1}{(1-p)^2}\int f_p \log \frac{f}{f_p}\\
& = - \frac{D(f_p\|f)}{(1-p)^2}\leq 0 
\end{align}
 where $D(\cdot\|\cdot)$ denotes the Kullback-Leibler divergence. Equality $D(f_p\|f)=0$ can hold only if $f=f_p$ a.e., which since $p\ne 1$ implies that $f$ is constant over some measurable subset $A\subset \R^n$, and is zero elsewhere. It follows that $h_p(X)> h_q(X)$ for any $p<q$ if $X$ is not uniformly distributed. Conversely, if $X$ is uniformly distributed over some measurable subset $A\subset \R^n$, its density can be written as $f(x)=1/\mathrm{vol}(A)$ for all $x\in A$ and $f(x)=0$ elsewhere. Then $h_p(X)=\frac{1}{1-p}\log \frac{\mathrm{vol}(A)}{\mathrm{vol}(A)^p}=\log \mathrm{vol}(A)$ is independent of $p$.
\end{proof}

\begin{Remark}
Notice the identity established in the proof:
\begin{equation}\label{identitydiff}
\frac{\partial}{\partial p} h_p(X) =   - \frac{D(f_p\|f)}{(1-p)^2}.
\end{equation}
A similar formula for discrete variables can be found in~\cite[\S~5.3]{BeckSchloegl93}.
\end{Remark}

\section{An Information Inequality}\label{iisec}

The Shannon entropy satisfies a fundamental ``information inequality''~\cite[Theorem~2.6.3]{CoverThomas06} from which many classical information-theoretic  inequalities can be derived. This can be written as
\begin{equation}\label{iih1}
h_1(X) \leq  -\E \log \phi(X) 
\end{equation}
for any density $\phi$, with equality if and only if $\phi=f$ a.e.\@  The following Theorem can be seen as the natural extension of the information inequality to R\'enyi entropies and is central in the following derivations of this paper.
J.F. Bercher~\cite{Bercher18} has pointed out to the author that it is similar to an inequality for discrete distributions established by Campbell~\cite{Campbell65} in the context of source coding (see also \cite{Bercher09}).

\begin{Theorem}[Information Inequality]\label{ii}
For any density $\phi$, 
\begin{equation}\label{iihp}
h_p(X) \leq -p' \log \E \bigl(\phi^{1/p'}(X) \bigr)
\end{equation}
with equality if and only if $\phi=f_p$ a.e.
\end{Theorem}
\noindent By letting $p\to 1$ one recovers the classical information inequality~\eqref{iih1} for the Shannon entropy. 

\begin{proof} By definition~\eqref{hpdef2},
\begin{align}
h_p(X)&=-p'\log\|f\|_p\\&=-p'\log\Bigl(\int (f^p\phi^{-1})\phi \Bigr)^{1/p} \\
&\leq -p' \log \int (f^p \phi^{-1})^{1/p} \phi \\
&= -p' \log \int f\phi^{1/p'}
\end{align}
where the inequality follows from Jensen's inequality applied to the function $x\mapsto x^{1/p}$, which is strictly concave if $p>1$ (that is, $p'>0$) and strictly convex if $p<1$ (that is, $p'<0$). Equality holds if and only if $f^p\phi^{-1}$ is constant a.e., which means that $\phi$ and $f^p$ are proportional a.e. Normalizing gives the announced condition $\phi=f_p$ a.e.
\end{proof}

\begin{Remark}
An alternate proof is obtained using Hölder's inequality or its reverse applied to $f$ and $\phi^{1/p'}$.
Notice that the equality case for $\phi=f_p$ gives
\begin{equation}\label{iiequalitycase}
h_p(X)= -p'\log\E\bigl( f_p^{1/p'}(X)\bigr) 
\end{equation}
as can be easily checked directly.  
\end{Remark}

The following conditional version of Theorem~\ref{ii} involves a more complicated relation for dependent variables.
\begin{Corollary}[Conditional Information Inequality]\label{condii}
For any two random variables $X,Y\in\R^n$,
\begin{equation}
-p' \log \E_Y \exp \bigl(- h_p(X|Y)/p' \bigr) \leq -p' \log \E \bigl(\phi^{1/p'}(X|Y) \bigr)
\end{equation}
where $h_p(X|y)$ denotes the R\'enyi entropy of $X$ knowing $Y=y$ and the expectation on the l.h.s. is taken over~$Y$ (the expectation in the r.h.s. is taken over $(X,Y)$). 

In particular, when $X$ and $Y$ are independent,
\begin{equation}\label{condiiindep}
h_p(X)\leq -p' \log \E \bigl(\phi^{1/p'}(X|Y) \bigr).
\end{equation}
with equality if and only if $\phi(x|y)$ does not depend on $y$ and equals $f_p(x)$ a.e.
\end{Corollary}
\begin{proof}
From~\eqref{iihp} for fixed $y$, one has $\E \bigl(\phi^{1/p'}(X|y) \bigr)\leq \exp \bigl(- h_p(X|y)/p' \bigr)$ for $p'>0$ ($p>1$) and the opposite inequality for $p'<0$ ($p<1$), with equality if and only if $\phi(x|y)=f_p(x|y)$ a.e.
Taking the expectation over $Y$ yields $\E \bigl(\phi^{1/p'}(X|Y) \bigr)\leq \E_Y\exp \bigl(- h_p(X|Y)/p' \bigr)$ for $p'>0$ ($p>1$) and the opposite inequality for $p'<0$ ($p<1$). The result follows by taking the logarithm and multiplying by $-p'$. When $X$ and $Y$ are independent, equality holds if and only if $\phi(x|y)=f_p(x)$ a.e. for all $y$.
\end{proof}

For the Shannon entropy, the difference between the two sides of the information inequality~\eqref{iih1} is the Kullback-Leibler divergence:
$$
D(f\|\phi)= \E \log\frac{f(X)}{\phi(X)} \geq 0
$$
which can also be noted $D(X\|Z)$ where $X\sim f$ and $Z\sim \phi$. It is known (and easy to check) that the divergence is invariant by reversible transformations $T$. This means that when $X=T(X^*)$ and $Z=T(Z^*)$, one has $D(X\|Z)=D(X^*\|Z^*)$. A natural extension to Rényi entropies can be obtained on the difference 
\begin{equation}
-p' \log \E \bigl(\phi^{1/p'}(X) \bigr) -h_p(X)  \geq 0
\end{equation}
between the two sides of the information inequality~\eqref{iihp}.
\begin{Theorem}[Transformational Invariance]\label{ti}
Let $T:\R^n\to\R^n$ be a diffeomorphism and
suppose that 
\begin{align}
X_p &= T(X^*_p)\\
Z &= T(Z^*) 
\end{align}
where $Z\sim \phi$ and $Z^*\sim \phi^*$. Then
\begin{equation}\label{tihp}
-p' \log \E \bigl(\phi^{1/p'}(X) \bigr) -h_p(X)=-p' \log \E \bigl({\phi^*}^{1/p'}(X^*) \bigr) -h_p(X^*).
\end{equation}
\end{Theorem}
\noindent Note that from~\eqref{hpdef2}, this identity can be rewritten as
\begin{equation}\label{invarianceidentity}
\frac{\E \bigl({\phi^*}^{1/p'}(X^*) \bigr)}{\|f^*\|_p} =\frac{ \E\bigl(\phi^{1/p'}(X)\bigr)  }{ \|f\|_p}.
\end{equation}

\begin{proof}
Proceed to prove~\eqref{invarianceidentity}.
Let $f,f^*$ be the respective densities of $X,X^*$ and recall that $X_p\sim f_p$ and $X^*_p\sim f^*_p$. By the transformation $T$ the densities are related by
\begin{align}
f^*_p(x^*) &= f_p(T(x^*)) |T'(x^*)|\\
\phi^*(x^*) &= \phi(T(x^*)) |T'(x^*)|\label{cv}
\end{align}
where $|T'|$ denotes the Jacobian determinant of $T$.
Using these relations and Definition~\ref{escort},
\begin{align}
\E \bigl({\phi^*}^{1/p'}(X^*) \bigr) / \|f^*\|_p&=\E \bigl(\phi^{1/p'}(T(X^*)) |T'(X^*)|^{1/p'}\bigr) / \|f^*\|_p\\
&=\int \phi^{1/p'}(T(x^*))  |T'(x^*)|^{1/p'} f^*(x^*)\d x^* / \|f^*\|_p\\
&= \int \phi^{1/p'}(T(x^*))  |T'(x^*)|^{1/p'} f^*_p(x^*)^{1/p}\d x^*\\
&= \int \phi^{1/p'}(T(x^*))   f_p(T(x^*))^{1/p} |T'(x^*)|\d x^*\\
&= \int \phi^{1/p'}(x)   f_p(x)^{1/p} \d x\\
&= \E\bigl(\phi^{1/p'}(X)\bigr)  / \|f\|_p.   \qedhere
\end{align}
\end{proof}

\begin{Remark}
The fact that $\phi$ is a density was not used in the proof of Theorem~\ref{ti}. Therefore~\eqref{tihp} holds more generally for any function $\phi$ satisfying~\eqref{cv}.
\end{Remark}

\section{First Version of the R\'enyi EPI}\label{repi1sec}

For two independent random variables $X$ and $Y$, the Shannon entropy power inequality can be expressed as follows~\cite{DemboCoverThomas91,Rioul11}: For any $0<\lambda<1$,
\begin{equation}
h(\sqrt{\lambda}X+ \sqrt{1-\lambda} Y) \geq \lambda h(X) + (1-\lambda) h(Y)
\end{equation}
with equality if and only if $X,Y$ are i.i.d.\@ normal\footnote{More precisely (given the translation invariance of the entropy) when the variables are normal with identical covariance matrices. Since it was assumed in this paper that all considered variables have zero mean, both statements are equivalent.\label{foot}}.
That is, the difference 
$h(\sqrt{\lambda}X+ \sqrt{1-\lambda} Y) - \lambda h(X) - (1-\lambda) h(Y)$
is minimum (zero) for i.i.d.\@ normal $X,Y$.
In this section, we study the natural generalization for Rényi entropies~\cite[Theorem~12]{DemboCoverThomas91}, namely that the quantity
\begin{equation}
h_r(\sqrt{\lambda}X+ \sqrt{1-\lambda} Y) - \lambda h_p(X) - (1-\lambda) h_q(Y) 
\end{equation}
is minimum for $X,Y$ i.i.d.\@ normal\footref{foot}.
Here the triple $(p,q,r)$ and its associated $\lambda$ satisfy  the following condition, which is used e.g., in Young's convolutional inequality.
\begin{Definition}[Exponent Triple]
An \emph{exponent triple} $(p,q,r)_\lambda$ has conjugates $p',q',r'$ of the same sign and such that
\begin{equation}
\frac{1}{p'}+\frac{1}{q'}=\frac{1}{r'}.
\end{equation}
The corresponding coefficient $\lambda\in(0,1)$ is defined by
\begin{equation}
\lambda= \frac{r'}{p'} = 1- \frac{r'}{q'}
\end{equation} 
\end{Definition}
\noindent In other words, the exponents $p,q,r$ are such that
\begin{equation}
\frac{1}{p}+\frac{1}{q}=1+\frac{1}{r} 
\end{equation}
and fulfill one the following two conditons:
\begin{center}
\begin{tabular}{|c|c|c|}
\toprule
$p,q,r>1$ &\raisebox{-1ex}{or}& $0<p,q,r<1$ \\
$p',q',r'>1$ && $p',q',r'<0$\\ 
\cmidrule(r){1-1}  \cmidrule(l){3-3}
$r'<p'$, $r'<q'$ && $|r'|<|p'|$, $|r'|<|q'|$\\
$r>p$, $r>q$ && $r<p$, $r<q$\\
\bottomrule 
\end{tabular} 
\end{center}
The key argument used in this section is the following. If $X\sim f$ and $Y\sim g$, then for the escort variables, $X_p\sim f_p$ and $Y_q\sim g_q$.  By normal transport (Lemma~\ref{gt}), one can write
\begin{align}
X_p &= T(X^*_p)\\
Y_q &= U(Y^*_q)
\end{align}
for two diffeomorphims $T$ and $U$, where $X^*,Y^*$ are, say, i.i.d.\@ standard normal $\mathcal{N}(0,\mathbf{I})$. 
(It follows that $X^*_p\sim \mathcal{N}(0,\mathbf{I}/p)$ and $Y^*_q\sim \mathcal{N}(0,\mathbf{I}/q)$.)
We then have the following straightforward extension of Theorem~\ref{ti}:

\begin{Lemma}[Transformational Invariance for Two Independent Variables]\label{titwo}
For a two-dimensional~$\phi(x,y)$,
\begin{equation}\label{invarianceidentitytwo} 
\begin{split}
-r' \log\E \big( \phi^{1/r'}(X,Y) \bigr) &- \lambda h_p(X) - (1-\lambda) h_q(Y)\\
&=
-r' \log\E \big( {\phi^*}^{1/r'}(X^*,Y^*) \bigr) - \lambda h_p(X^*) - (1-\lambda) h_q(Y^*) 
\end{split}
\end{equation}
where 
\begin{equation}
{\phi^*}(x^*,y^*)=\phi(T(x^*),U(y^*)) |T'(x^*)|^\lambda |U'(y^*)|^{1-\lambda}.
\end{equation}
\end{Lemma}

\begin{proof}
From~\eqref{hpdef2} and the definition of $\lambda$, \eqref{invarianceidentitytwo} can be rewritten as
\begin{equation}
\frac{\E \bigl({\phi^*}^{1/r'}(X^*,Y^*) \bigr) }{ \|f^*\|_p \|g^*\|_q}=
 \frac{\E\bigl(\phi^{1/r'}(X,Y)\bigr)  }{ \|f\|_p \|g\|_q}.
\end{equation}
By the transformations $T$ and $U$ the densities of the escort variables are related by $f^*_p(x^*) = f_p(T(x^*)) |T'(x^*)|$ and $g^*_q(y^*) = g_q(U(y^*)) |U'(y^*)|$.
Now by the same calculation as in the proof of Theorem~\ref{ti},
\begin{align}
\frac{\E \bigl({\phi^*}^{1/r'}(X^*,Y^*) \bigr) }{ \|f^*\|_p \|g^*\|_q}
&=\frac{\E \bigl(\phi^{1/r'}(T(X^*),U(Y^*)) |T'(X^*)|^{1/p'} |U'(Y^*)|^{1/q'}\bigr) }{ \|f^*\|_p \|g^*\|_q} \\
&=\frac{\int \phi^{1/r'}\bigl(T(x^*),U(y^*)\bigr)  |T'(x^*)|^{1/p'} |U'(y^*)|^{1/q'} f^*(x^*) g^*(y^*)\d x^*\d y^* }{ \|f^*\|_p \|g^*\|_q}\\
&= \int \phi^{1/r'}\bigl(T(x^*),U(y^*)\bigr)  |T'(x^*)|^{1/p'} |U'(y^*)|^{1/q'} f_p^*(x^*)^{1/p} g_q^*(y^*)^{1/q}\d x^*\d y^* \\
&= \int \phi^{1/r'}\bigl(T(x^*),U(y^*)\bigr)  |T'(x^*)| |U'(y^*)| f_p(T(x^*))^{1/p} g_q(U(y^*))^{1/q}\d x^*\d y^*\\
&= \int \phi^{1/r'}(x,y)  f_p(x)^{1/p} g_q(y)^{1/q}\d x\d y\\
&= \frac{\E\bigl(\phi^{1/r'}(X,Y)\bigr)  }{ \|f\|_p \|g\|_q}  .\qedhere
\end{align}
\end{proof}

\begin{Lemma}\label{firststep}
Let $\phi$ be the density of $\sqrt{\lambda}X+\sqrt{1-\lambda}Y$. Then
\begin{equation}\label{firststepineq}
\begin{split}
h_r(\sqrt{\lambda}X&+\sqrt{1-\lambda}Y) - \lambda h_p(X) -(1-\lambda)h_q(Y)\\
&\geq -r' \log \E \Bigl\{\Bigl( \phi_r (\sqrt{\lambda}T(X^*)+\sqrt{1-\lambda}U(Y^*)) \cdot |\lambda T'(X^*) + (1-\lambda) U'(Y^*)| \Bigr)^{1/r'} \Bigr\}\\
&\qquad - \lambda h_p(X^*) -(1-\lambda)h_q(Y^*).
\end{split}
\end{equation}
\end{Lemma}

\begin{proof}
By the equality case of Theorem~\ref{ii} (see~\eqref{iiequalitycase}), one has
\begin{equation}
h_r(\sqrt{\lambda}X+\sqrt{1-\lambda}Y) = - r' \log \E \phi_r^{1/r'}(\sqrt{\lambda}X+\sqrt{1-\lambda}Y).
\end{equation}
Now by Lemma~\ref{titwo} applied to $\phi(x,y)=\phi_r(\sqrt{\lambda}x+\sqrt{1-\lambda}y)$,
we have $\phi^*(x^*,y^*)=\phi_r(\sqrt{\lambda}T(x^*)+\sqrt{1-\lambda}U(y^*)) |T'(x^*)|^\lambda |U'(y^*)|^{1-\lambda}$,
and, therefore,
\begin{equation}
\begin{split}
h_r(\sqrt{\lambda}X&+\sqrt{1-\lambda}Y) - \lambda h_p(X) - (1-\lambda) h_q(Y)\\
&=
-r' \log \E \Bigl\{\Bigl( \phi_r (\sqrt{\lambda}T(X^*)+\sqrt{1-\lambda}U(Y^*)) \cdot |T'(X^*)|^\lambda |U'(Y^*)|^{1-\lambda} \Bigr)^{1/r'} \Bigr\}\\ &\qquad- \lambda h_p(X^*) - (1-\lambda) h_q(Y^*). 
\end{split}
\end{equation}
Since from Lemma~\ref{gt}, $T$ and $U$ can be chosen such that $T'$ and $U'$ are (lower) triangular with positive diagonal elements, it follows easily from the arithmetic-geometric mean inequality that 
\begin{equation}\label{amgm}
|T'(X^*)|^\lambda |U'(Y^*)|^{1-\lambda}\leq |\lambda T'(X^*) + (1-\lambda) U'(Y^*)|. 
\end{equation}
 The result follows at once (for either positive or negative $r'$).
\end{proof}

We can now use the normal rotation Lemma~\ref{gr} to conclude by proving the following
\begin{Theorem}[R\'enyi EPI~\cite{DemboCoverThomas91}]\label{repi1}
For independent $X,Y$ and exponent triple $(p,q,r)_\lambda$,
\begin{equation}\label{repi1ineq}
h_r(\sqrt{\lambda}X+ \sqrt{1-\lambda} Y) - \lambda h_p(X) - (1-\lambda) h_q(Y) \geq \frac{n}{2} r' \Bigl( \frac{\log r}{r} - \frac{\log p}{p} - \frac{\log q}{q} \Bigr)
\end{equation}
with equality if and only if $X,Y$ are i.i.d.\@ normal\footref{foot}.
\end{Theorem}

\begin{proof}
If $X,Y$ are i.i.d.\@ normal, then $\sqrt{\lambda}X+ \sqrt{1-\lambda} Y$ is also identically distributed as $X$ and $Y$, and
from Lemma~\ref{gausshp}, it is immediate to check that equality holds (irrespective of their covariances). Therefore, inequality~\eqref{repi1ineq} is equivalent to 
\begin{equation}\label{repi1equiv}
h_r(\sqrt{\lambda}X+ \sqrt{1-\lambda} Y) - \lambda h_p(X) - (1-\lambda) h_q(Y) \geq
h_r(\sqrt{\lambda}X^*+ \sqrt{1-\lambda} Y^*) - \lambda h_p(X^*) - (1-\lambda) h_q(Y^*)  
\end{equation}
where $X^*$, $Y^*$ are, say, i.i.d.\@ standard normal $\mathcal{N}(0,\mathbf{I})$. 

To prove~\eqref{repi1equiv}, consider the normal rotation of Lemma~\ref{gr} and write $X^*,Y^*$ in terms of $\tilde{X},\tilde{Y}$ using~\eqref{invrot} in the first term of the r.h.s. of~\eqref{firststepineq} (Lemma~\ref{firststep}).
One obtains:
\begin{equation}\label{afterrotation}
\begin{split}
h_r(\sqrt{\lambda}X&+\sqrt{1-\lambda}Y) - \lambda h_p(X) -(1-\lambda)h_q(Y)\\
&\geq -r' \log \E \Bigl\{\bigl(   \psi(\tilde{X}|\tilde{Y})   \bigr)^{1/r'} \Bigr\} - \lambda h_p(X^*) -(1-\lambda)h_q(Y^*),
\end{split}
\end{equation}
where 
\begin{equation}\label{psi}
\begin{split}
\psi(\tilde{x}|\tilde{y}) =  &\phi_r (\sqrt{\lambda}T(\sqrt{\lambda}\tilde{x}-\sqrt{1-\lambda}\tilde{y})+\sqrt{1-\lambda}U(\sqrt{1-\lambda}\tilde{x}+\sqrt{\lambda}\tilde{y})) \\&\times |\lambda T'(\sqrt{\lambda}\tilde{x}-\sqrt{1-\lambda}\tilde{y}) + (1-\lambda) U'(\sqrt{1-\lambda}\tilde{x}+\sqrt{\lambda}\tilde{y})| .
\end{split}
\end{equation}
Making the change of variable $z=\sqrt{\lambda}T(\sqrt{\lambda}\tilde{x}-\sqrt{1-\lambda}\tilde{y})+\sqrt{1-\lambda}U(\sqrt{1-\lambda}\tilde{x}+\sqrt{\lambda}\tilde{y})$, one obtains 
\begin{equation}
\int \psi(\tilde{x}|\tilde{y}) \d \tilde{x} = \int \phi_r(z)\d z =1,
\end{equation}
since $\phi_r$ is a density. Hence $\psi(\tilde{x}|\tilde{y})$ is also a density in $\tilde{x}$ for fixed $\tilde{y}$. Now since by Lemma~\ref{gr}, $\tilde{X}$ and $\tilde{Y}$ are independent, by the conditional information inequality~\eqref{condiiindep} of Corollary~\ref{condii}, one has
\begin{equation}\label{condpsi}
-r' \log \E \Bigl\{\bigl(   \psi(\tilde{X}|\tilde{Y})   \bigr)^{1/r'} \Bigr\} \geq  h_r(\tilde{X})=h_r(\sqrt{\lambda}X^*+\sqrt{1-\lambda}Y^*).
\end{equation}
Combining with~\eqref{afterrotation} yields the announced inequality~\eqref{repi1equiv}.

It remains to settle the equality case in~\eqref{repi1equiv}. From the above proof, equality holds in~\eqref{repi1equiv} if and only if both~\eqref{amgm} and~\eqref{condpsi} are equalities.
Equality in~\eqref{amgm} holds if and only if for all $i=1,2,\ldots,n$,
\begin{equation}
\frac{\partial T_i}{\partial x_i}(X^*) =  \frac{\partial U_i}{\partial y_i}(Y^*) \text{ a.s.}
\end{equation}
Since $X^*$ and $Y^*$ are independent normal variables, this implies that $\frac{\partial T}{\partial x_i}$ and $\frac{\partial U}{\partial y_i}$ are constant and equal. In particular the Jacobian $|\lambda T'(\sqrt{\lambda}\tilde{x}-\sqrt{1-\lambda}\tilde{y}) + (1-\lambda) U'(\sqrt{1-\lambda}\tilde{x}+\sqrt{\lambda}\tilde{y})| $ in~\eqref{psi} is constant. 

From Corollary~\ref{condii} equality in~\eqref{condpsi} holds if and only if $ \psi(\tilde{x}|\tilde{y})$ does not depend on $\tilde{y}$, which implies that $\sqrt{\lambda}T(\sqrt{\lambda}\tilde{x}-\sqrt{1-\lambda}\tilde{y})+\sqrt{1-\lambda}U(\sqrt{1-\lambda}\tilde{x}+\sqrt{\lambda}\tilde{y})$ does not depend on the value of $\tilde{y}$. Taking derivatives with respect to $y_j$ for all $j=1,2,\ldots,n$,
\begin{equation}
  -\sqrt{\lambda}\sqrt{1-\lambda} \frac{\partial T_i}{\partial x_j}(\sqrt{\lambda}\tilde{X}-\sqrt{1-\lambda}\tilde{Y}) + \sqrt{\lambda}\sqrt{1-\lambda} \frac{\partial U_i}{\partial x_j}(\sqrt{1-\lambda}\tilde{X}+\sqrt{\lambda}\tilde{Y})=0
\end{equation}
which implies 
\begin{equation}
\frac{\partial T_i}{\partial x_j}(X^*) =  \frac{\partial U_i}{\partial y_j}(Y^*) \text{ a.s.}
\end{equation}
for all $i,j=1,2,\ldots,n$. Therefore, $T$ and $U$ are linear transformations, equal up to an additive constant (equal to $0$ since all variables are assumed of zero mean). It follows that $X_p = T(X^*_p)$ and $Y_q = U(Y^*_q)$ are normal with respective distributions $X_p\sim \mathcal{N}(0,\mathbf{K}/p)$ and $Y_q\sim \mathcal{N}(0,\mathbf{K}/q)$. Hence $X$ and $Y$ are i.i.d.\@ normal $\mathcal{N}(0,\mathbf{K})$.
\end{proof}

A straightforward generalization to several independent variables is the following
\begin{Corollary}[R\'enyi EPI for Several Variables]\label{repi1m}
Let $r_1,r_2,\ldots,r_m,r$ be exponents those conjugates $r'_1,r'_2,\ldots,r'_m,r'$ are of the same sign and satisfy
\begin{equation}
\sum_{i=1}^m\frac{1}{r'_i}=\frac{1}{r'}
\end{equation}
and let $\lambda_1,\lambda_2,\ldots,\lambda_m$ be defined by
\begin{equation}
\lambda_i =  \frac{r'}{r'_i}\qquad (i=1,2,\ldots,m).
\end{equation}
Then for independent $X_1,X_2,\ldots,X_m$,
\begin{equation}\label{repi1mineq}
h_r\Bigl( \sum_{i=1}^m\sqrt{\lambda_i}X_i\Bigr) - \sum_{i=1}^m \lambda_i h_{r_i}(X_i)
\geq \frac{n}{2} r' \Bigl( \frac{\log r}{r}-\sum_{i=1}^m\frac{\log r_i}{r_i}\Bigr)
\end{equation}
with equality if and only if the $X_i$ are i.i.d.\@ normal\footref{foot}.
\end{Corollary}

\begin{proof}
By induction on $m$:  The result for $m=2$ is Theorem~\ref{repi1}. Suppose the result satisfied at order $m-1$ and let $Y_m=\sum_{i=1}^{m-1}\sqrt{\lambda_i}X_i/\sqrt{1-\lambda_m}$ and $s_m$ be such that $\frac{1}{s'_m}=\sum_{i=1}^{m-1} \frac{1}{r'_i}$.
Notice that $\frac{1}{r'}=\frac{1}{r'_m}+\frac{1}{s'_m} = \frac{\lambda_m}{r'}+\frac{1}{s'_m}$, hence $r'=(1-\lambda_m) s'_m$.
By Theorem~\ref{repi1}, $h_r\Bigl( \sum_{i=1}^m\sqrt{\lambda_i}X_i\Bigr)=h_r(\sqrt{\lambda_m}X_m+\sqrt{1-\lambda_m}Y_m)
\geq \lambda_m h_{r_m}(X_m) + (1-\lambda_m) h_{s_m}(Y_m) +\frac{n}{2}r' \bigl(\frac{\log r}{r}-\frac{\log r_m}{r_m}-\frac{\log s_m}{s_m} \bigr)$ with equality if and only if $X_m,Y_m$ are i.i.d.\@ normal. Now by the induction hypothesis, $h_{s_m}(Y_m)\geq \sum_{i=1}^{m-1} \frac{\lambda_i}{1-\lambda_m}h_{r_i}(X_i) + \frac{n}{2} s'_m \bigl(\frac{\log s_m}{s_m}-\sum_{i=1}^{m-1}\frac{\log r_i}{r_i} \bigr)$ with equality if and only if the $X_i$ ($i=1,2,\ldots,m-1$)---and hence $Y_m$---are i.i.d.\@ normal.
The result at order $m$ follows by combining the two inequalities since $(1-\lambda_m) s'_m = r'$.
\end{proof}

\section{Recent Versions of the Rényi EPI}\label{repi2sec}

\begin{Definition}[Rényi Entropy Power~\cite{BobkovChistyakov15}] \label{RenyiEntropyPower}
The \emph{Rényi entropy power} of order $r$ is defined by
\begin{equation}
N_r(X)= e^{2h_r(X)/n}.
\end{equation}
\end{Definition}
Up to a multiplicative constant, $N_r(X)$ is the (average) power of a white normal variable having the same Rényi entropy as $X$---hence the name ``entropy power''. In fact, if $X^*\sim\mathcal{N}(0,\sigma^2)$ has the same Rényi entropy $h_r(X^*)=h_r(X)$, then by Lemma~\ref{gausshp},
\begin{equation}
\sigma^2 =  \frac{e^{2h_r(X)/n}}{2\pi r^{r'/r}}.
\end{equation}
The Renyi entropy power enjoys the same scaling property as for the usual power: By Lemma~\ref{scaling}, for any $a\in\R$,
\begin{equation}\label{repiscale}
N_r(aX)=a^2 N_r(X).
\end{equation}

For independent $X_1,X_2,\ldots,X_m$, R\'enyi entropy power inequalities take either the form~\cite{BobkovChistyakov15,RamSason16}
\begin{equation}\label{repic}
N_r\Bigl( \sum_{i=1}^m X_i\Bigr) \geq c  \sum_{i=1}^m N_r(X_i)
\end{equation}
for some positive constant $c$, or the form~\cite{BobkovMarsiglietti17,Li18,MarsigliettiMelbourne18}
\begin{equation}\label{repialpha}
{N_r^{\vphantom{2}}}^\alpha\Bigl( \sum_{i=1}^m X_i\Bigr) \geq  \sum_{i=1}^m {N_r^{\vphantom{2}}}^\alpha(X_i)
\end{equation}
for some positive exponent $\alpha$. The constants $c$ and $\alpha$ may depend on the order $r$, the number $m$ of variables and the dimension~$n$. What is desired is:
\begin{itemize}
\item
a maximum possible value of $c$ in~\eqref{repic} since the inequality is automatically satisfied for all positive constants $c'<c$.
\item 
a minimum possible value of $\alpha$ in~\eqref{repialpha} since the inequality is automatically satisfied for all positive exponents $\alpha'>\alpha$;
in fact, since~\eqref{repialpha} is homogeneous by scaling the variables $X_i\mapsto aX_i$ as in~\eqref{repiscale}, one may suppose without loss of generality that the r.h.s. of~\eqref{repialpha} is $=1$; then $N_r(X_i)<1$ hence ${N_r^{\vphantom{2}}}^{\alpha'}(X_i)<{N_r^{\vphantom{2}}}^{\alpha}(X_i)$ for all $i$ and $1=\sqrt[\alpha]{\sum_{i=1}^m {N_r^{\vphantom{2}}}^{\alpha}(X_i)}\geq \sqrt[\alpha']{\sum_{i=1}^m {N_r^{\vphantom{2}}}^{\alpha'}(X_i)}$.
\end{itemize}
The following useful characterization, which generalizes~\cite[Lemma~2.1]{Li18}, makes the link between the various versions~\eqref{repi1mineq},~\eqref{repic},~\eqref{repialpha} of the Rényi entropy power inequality.
\begin{Lemma}\label{charact}
For independent $X_1,X_2,\ldots,X_m$, the Rényi EPI in the general form 
\begin{equation}\label{repig}
{N_r^{\vphantom{2}}}^\alpha\Bigl( \sum_{i=1}^m X_i\Bigr) \geq  c\sum_{i=1}^m {N_r^{\vphantom{2}}}^\alpha(X_i)
\end{equation}
for some constant $c>0$ and exponent $\alpha>0$ is equivalent to the following inequality
\begin{equation}\label{repi1bis}
h_r\Bigl( \sum_{i=1}^m\sqrt{\lambda_i}X_i\Bigr) - \sum_{i=1}^m \lambda_i h_{r}(X_i)
\geq \frac{n}{2} \Bigl(  \frac{\log c}{\alpha} + \bigl(\frac{1}{\alpha}-1\bigr) H(\lambda)  \Bigr)
\end{equation}
for any positive $\lambda_1,\lambda_2,\ldots,\lambda_m$ such that $\sum_{i=1}^m \lambda_i=1$, where
$H(\lambda)> 0$ denotes the discrete entropy
\begin{equation}
H(\lambda) = \sum_{i=1}^m \lambda_i\log \frac{1}{\lambda_i} >0.
\end{equation}
\end{Lemma}
\begin{proof}
Suppose~\eqref{repig} holds. Then
\begin{align}
h_r\Bigl( \sum_{i=1}^m\sqrt{\lambda_i}X_i\Bigr)  &=\frac{n}{2\alpha}\log {N_r^{\vphantom{2}}}^\alpha\Bigl( \sum_{i=1}^m\sqrt{\lambda_i}X_i\Bigr)\\
&\geq \frac{n}{2\alpha}\log\sum_{i=1}^m {N_r^{\vphantom{2}}}^\alpha(\sqrt{\lambda_i}X_i) + \frac{n}{2\alpha}\log c\\
&=\frac{n}{2\alpha}\log\sum_{i=1}^m \lambda_i^\alpha {N_r^{\vphantom{2}}}^\alpha(X_i) + \frac{n}{2\alpha}\log c   \label{a}\\
&\geq \frac{n}{2\alpha}\sum_{i=1}^m \lambda_i \log\bigl( \lambda_i^{\alpha-1} {N_r^{\vphantom{2}}}^\alpha(X_i)\bigr) + \frac{n}{2\alpha}\log c  
\label{b}\\
&= \sum_{i=1}^m \lambda_i h_{r}(X_i) +\frac{n(\alpha-1)}{2\alpha}\sum_{i=1}^m \lambda_i\log {\lambda_i}
+ \frac{n}{2\alpha}\log c 
\end{align}
where the scaling property~\eqref{repiscale} is used in~\eqref{a} and the concavity of the logarithm is used in~\eqref{b}. Conversely, suppose that~\eqref{repi1bis} is satisfied for all $\lambda_i>0$ such that $\sum_{i=1}^m \lambda_i=1$. Set $\lambda_i = {N_r^{\vphantom{2}}}^\alpha(X_i)/ \sum_{i=1}^m {N_r^{\vphantom{2}}}^\alpha(X_i)$.
 Then
\begin{align}
{N_r^{\vphantom{2}}}^\alpha\Bigl( \sum_{i=1}^m X_i\Bigr) 
&= \exp \frac{2\alpha}{n} h_r\Bigl( \sum_{i=1}^m\sqrt{\lambda_i}\frac{X_i}{\sqrt{\lambda_i}}\Bigr)\\
&\geq \exp \frac{2\alpha}{n} \sum_{i=1}^m \lambda_i h_{r}\Bigl(\frac{X_i}{\sqrt{\lambda_i}}\Bigr)
\cdot c \cdot \exp (1-\alpha) \sum_{i=1}^m \lambda_i\log \frac{1}{\lambda_i}\\
&= c \prod_{i=1}^m  \Biggl({N_r^{\vphantom{2}}}^\alpha\Bigl(\frac{X_i}{\sqrt{\lambda_i}}\Bigr) \lambda_i^{\alpha-1} \Biggr)^{\lambda_i} \\
&= c \prod_{i=1}^m  \Bigl({N_r^{\vphantom{2}}}^\alpha(X_i)
\lambda_i^{-1} \Bigr)^{\lambda_i} \\
&=c\Bigl(\sum_{i=1}^m {N_r^{\vphantom{2}}}^\alpha(X_i)\Bigr)^{\sum_{i=1}^m \lambda_i}\\
&=c\sum_{i=1}^m {N_r^{\vphantom{2}}}^\alpha(X_i).\qedhere
\end{align}
\end{proof}

\subsection{Rényi Entropy Power Inequalities for Orders >1}

From Lemma~\ref{charact} and Corollary~\ref{repi1m} it is easy to recover known Rényi EPIs and obtain new ones for orders $r>1$.
In fact, if $r>1$ then $r'>0$ and all $r'_i$ are positive and $>r'$. Therefore all $r_i$ are $<r$ and by monotonicity (Lemma~\ref{mono}), 
\begin{equation}\label{monori}
h_{r_i}(X_i)\geq h_r(X_i) \qquad (i=1,2,\ldots,m).  
\end{equation}
Plugging this into~\eqref{repi1mineq} one obtains
\begin{equation}\label{repi2mineq}
h_r\Bigl( \sum_{i=1}^m\sqrt{\lambda_i}X_i\Bigr) - \sum_{i=1}^m \lambda_i h_{r}(X_i)
\geq \frac{n}{2} r' \Bigl( \frac{\log r}{r}-\sum_{i=1}^m\frac{\log r_i}{r_i}\Bigr)
\end{equation}
where $\lambda_i=r'/r'_i$ for $i=1,2,\ldots,m$. For future reference define\footnote{The absolute value $|r'|$ is needed in the next subsection where $r'$ will be negative.}
\begin{align}
A(\lambda)&=|r'| \Bigl( \frac{\log r}{r}-\sum_{i=1}^m\frac{\log r_i}{r_i}\Bigr)\\
&=|r'| \Bigl( 
\sum_{i=1}^m (1-\frac{\lambda_i}{r'})\log(1-\frac{\lambda_i}{r'})
-(1-\frac{1}{r'})\log(1-\frac{1}{r'})
\Bigr).
\label{Alambda}
\end{align}
This function is strictly convex in $\lambda=(\lambda_1,\lambda_2,\ldots,\lambda_m)$ because 
$x\mapsto (1-x/r')\log(1-x/r')$ is strictly convex. Note that $A(\lambda)$ vanishes in the limiting cases where $\lambda$ tends to one of the standard unit vectors $(1,0,\ldots,0)$, $(0,1,0,\ldots,0)$, \ldots, $(0,0,\ldots,0,1)$ and since every $\lambda$ is a convex combination of these vectors and $A(\lambda)$ is strictly convex, one has $A(\lambda)<0$.

\begin{Theorem}[Ram and Sason~\cite{RamSason16}]\label{thmrepic}
The Rényi EPI~\eqref{repic} holds for $r>1$ and $c=r^{r'/r}\bigl(1-\frac{1}{mr'}\bigr)^{mr'-1}$.
\end{Theorem}

\begin{proof}
By Lemma~\ref{charact} for $\alpha=1$ we only need to check that the r.h.s. of~\eqref{repi2mineq} is greater than $\frac{n}{2}\log c$ for any choice of the $\lambda_i$'s, that is, for any choice of exponents $r_i$ such that $\sum_{i=1}^m \frac{1}{r'_i} = \frac{1}{r'}$. Thus~\eqref{repic} will hold for 
$\log c = \min_{\lambda} A(\lambda)$. Now by the log-sum inequality~\cite[Theorem~2.7.1]{CoverThomas06},
\begin{equation}
\sum_{i=1}^m\frac{1}{r_i}\log \frac{1}{r_i} \geq  \Bigl(\sum_{i=1}^m\frac{1}{r_i}\Bigr) \log \frac{\sum_{i=1}^m\frac{1}{r_i}}{m} = (m-1/r')\log \frac{m-1/r'}{m}
\end{equation}
with equality if and only if all $r_i$ are equal, that is, the $\lambda_i$ are equal to $1/m$. Thus 
$\min_\lambda A(\lambda) =  r'  \Bigl( \frac{\log r}{r}+ (m-1/r')\log \frac{m-1/r'}{m} \Bigr)$ which yields $c=r^{r'/r}\bigl(1-\frac{1}{mr'}\bigr)^{mr'-1}$.
\end{proof}
An alternate proof is to argue that $A(\lambda)$ is convex and symmetrical in $\lambda=(\lambda_1,\lambda_2,\ldots,\lambda_m)$ and is, therefore, minimized when all $\lambda_i$ are equal.

\begin{Remark}
The above constant $c$ is certainly not optimal since equality in~\eqref{monori} holds if and only if the $X_i$ are uniformly distributed (Lemma~\ref{mono}) while equality in~\eqref{repi1mineq} holds if and only if the $X_i$ are identically normally distributed (Corollary~\ref{repi1m}). Ram and Sason~\cite{RamSason16} tightened~\eqref{repic} further using optimization techniques, resulting in a constant that depends on the relative values of the entropy powers themselves. 
\end{Remark}

\begin{Remark}\label{logc}
It can be noted that $\log c= r' \frac{\log r}{r} +(mr'-1) \log \bigl(1-\frac{1}{mr'}\bigr)<0$ decreases (and tends to $r' \frac{\log r}{r} -1$) as $m$ increases; in fact $\frac{\partial \log c}{\partial m} = r' \log \bigl(1-\frac{1}{mr'}\bigr) +\frac{mr'}{r'm^2} < r'(-\frac{1}{mr'})+\frac{1}{m}=0$. 
Thus, a universal constant independent of $m$ is obtained by taking
\begin{equation}
c= \inf_m \;r^{r'/r}\bigl(1-\frac{1}{mr'}\bigr)^{mr'-1} = r^{r'/r}\lim_{m\to\infty}\bigl(1-\frac{1}{mr'}\bigr)^{mr'-1}=\frac{r^{r'/r}}{e}
\end{equation}
This was the constant established by Bobkov and Chistyakov~\cite{BobkovChistyakov15}.
\end{Remark}

\begin{Theorem}\label{thmrepialpha}
The Rényi EPI~\eqref{repialpha} holds for $r>1$ and 
$\alpha=\Bigl(1+{r' \frac{\log_2 r}{r} +(2r'-1) \log_2 \bigl(1-\frac{1}{2r'}\bigr)}  \Bigr)^{-1}$
and this value of $\alpha$ cannot be improved using the method of this paper by making it depend on $m$.
\end{Theorem}

\begin{proof}
By Lemma~\ref{charact} for $c=1$ we only need to check that the r.h.s. of~\eqref{repi2mineq} is greater than $\frac{n}{2}(1/\alpha-1)H(\lambda)$ for any choice of the $\lambda_i$'s, that is, for any choice of exponents $r_i$ such that $\sum_{i=1}^m \frac{1}{r'_i} = \frac{1}{r'}$. Thus~\eqref{repialpha} will hold for 
$\frac{1}{\alpha}-1 = \min_\lambda \frac{A(\lambda)}{H(\lambda)}$.
By the proof of the preceding theorem, the numerator is minimized when all $\lambda_i$ are equal and this also maximizes the entropy $=\log m$ in the denominator.  However one cannot conclude yet since the minimum in the numerator is negative. 

A stationary point is easily obtained by the Lagrangian method which implies that $\frac{\partial}{\partial \lambda_i} \frac{A(\lambda)}{H(\lambda)}$ is constant independent of $i$. This gives that $\frac{A(\lambda)}{H(\lambda)} \log \lambda_i - \log (1-\lambda_i/r')$ is constant, hence a stationary point is obtained when all $\lambda_i$ are equal (to $1/m$) and the corresponding value of $A(\lambda)/H(\lambda)$ is $\Bigl(r' \frac{\log r}{r} +(mr'-1) \log \bigl(1-\frac{1}{mr'}\bigr)\Bigr)/\log m$.

However, the boundary of the domain of $A(\lambda)/H(\lambda)$ is the simplex $\{\sum_i \lambda_i=1, \lambda_i\geq 0\}$ where on each vertex joining two standard unit vectors, $A(\lambda)/H(\lambda)$ has the same expression as for $m=2$. Now Li~\cite{Li18} has shown\footnote{This can also be easily proved using~\cite[Lemma~8]{MarsigliettiMelbourne18}.} that for $m=2$, the minimum is obtained when $\lambda=(1/2,1/2)$. The correponsding value of  $A(\lambda)/H(\lambda)$ is $\Bigl(r' \frac{\log r}{r} +(2r'-1) \log \bigl(1-\frac{1}{2r'}\bigr)\Bigr)/\log 2$, which is easily seen to be less than $\Bigl(r' \frac{\log r}{r} +(mr'-1) \log \bigl(1-\frac{1}{mr'}\bigr)\Bigr)/\log m$ for any $m\geq 2$.

Therefore, the minimum of $A(\lambda)/H(\lambda)$ is attained at the boundary when all $\lambda_i$ are zero except two of them equal to $1/2$.
This gives $\frac{1}{\alpha}-1=\Bigl(r' \frac{\log r}{r} +(2r'-1) \log \bigl(1-\frac{1}{2r'}\bigr)\Bigr)/\log 2$.
\end{proof}

\begin{Remark}
The case $m=2$ yields $\alpha=\frac{r-1}{(r+1)\log_2(r+1)-r\log_2 r -2}$ which was found by Li~\cite{Li18} who remarked that this value of $\alpha$ is strictly smaller (better) than the value $\alpha=\frac{r+1}{2}$ obtained by Bobkov and Marsiglietti~\cite{BobkovMarsiglietti17}.

Interestingly, for $m>2$ the exponent of Theorem~\ref{thmrepialpha} cannot be improved by this method. In fact, in the above proof it easily seen that $\Bigl(r' \frac{\log r}{r} +(mr'-1) \log \bigl(1-\frac{1}{mr'}\bigr)\Bigr)/\log m$ is negative and increases toward $0$ as $m$ increases. Therefore, the exponent $\alpha$ cannot be decreased (improved) as $m$ increases. 
\end{Remark}

The above value of $\alpha$ is $>1$. However, using the same method it is easy to obtain Rényi EPIs with exponent values $\alpha<1$. This is given by the following Theorem.

\begin{Theorem}\label{thmrepig}
The Rényi EPI~\eqref{repig} holds for $r>1$, $0<\alpha<1$ with $c=\Bigl(m\;r^{r'/r}\bigl(1-\frac{1}{mr'}\bigr)^{mr'-1}\Bigr)^\alpha / m$.
\end{Theorem}
\begin{proof}
By Lemma~\ref{charact} we only need to check that the r.h.s. of~\eqref{repi2mineq} is greater than $\frac{n}{2}\bigl((\log c)/\alpha+(1/\alpha-1)H(\lambda)\bigr)$, that is, $A(\lambda)\geq (\log c)/\alpha+(1/\alpha-1)H(\lambda)$ for any choice of the $\lambda_i$'s, that is, for any choice of exponents $r_i$ such that $\sum_{i=1}^m \frac{1}{r'_i} = \frac{1}{r'}$. Thus for a given $0<\alpha<1$, \eqref{repig} will hold for $\log c = \min_\lambda \alpha A(\lambda) - (1-\alpha) H(\lambda)$. From the preceding proofs (since both $A(\lambda)$ and $-H(\lambda)$ are convex functions of $\lambda$) the minimum is attained when all $\lambda_i$ are equal. This gives $\log c =\alpha \Bigl(r' \frac{\log r}{r} +(mr'-1) \log \bigl(1-\frac{1}{mr'}\bigr)\Bigr)-(1-\alpha) \log m$.
\end{proof}

 
\subsection{Rényi Entropy Power Inequalities for Orders <1 and Log-Concave Densities}

If $r<1$ then $r'<0$ and all $r'_i$ are negative and $<r'$. Therefore all $r_i$ are $>r$ 
and by monotonicity (Lemma~\ref{mono}), the opposite inequality of~\eqref{monori} holds and the method of the preceding subsection fails. For log-concave densities, however,~\eqref{monori} can be replaced by a similar inequality in the right direction.

\begin{Definition}[Log-Concave Density]
A density $f$ is log-concave if $\log f$ is concave in it support, i.e., for all $0<\mu<1$,
\begin{equation}\label{logconcave}
f(x)^\mu f(y)^{1-\mu} \leq  f(\mu x+ (1-\mu) y).
\end{equation}
\end{Definition}

\begin{Lemma}\label{logconcaveentropyconcave}
If $X$ has a log-concave density, then $h_p(pX)-ph_p(X)=n\log p + (1-p) h_p(X)$ is concave in~$p$.
\end{Lemma}
As noted below in Corollary~\ref{varentropy}, this is essentially a result obtained by Fradelizi, Madiman and Wang~\cite{FradeliziMadimanWang16}. The following alternate proof uses the transport properties seen in Section~\ref{repi1sec}.
\begin{proof}
Define $r=\lambda p + (1-\lambda) q$ where $0<\lambda<1$. By Lemma~\ref{gt} there exists two diffeomorphisms $T,U$ such that one can write $pX_p=T(X^*)$ and $qX_q=U(X^*)$. Then $X^*$ has density
\begin{equation}\label{densitylambda}
\tfrac{1}{p^n} f_p\bigl(\tfrac{T(x^*)}{p}\bigr) |T'(x^*)| = 
\tfrac{1}{q^n} f_q\bigl(\tfrac{U(x^*)}{q}\bigr) |U'(x^*)| =
\tfrac{1}{p^{\lambda n}q^{(1-\lambda)n}} f_p\bigl(\tfrac{T(x^*)}{p}\bigr)^\lambda f_q\bigl(\tfrac{U(x^*)}{q}\bigr)^{1-\lambda} |T'(x^*)|^\lambda|U'(x^*)|^{1-\lambda}.
\end{equation}
Now, by log-concavity~\eqref{logconcave} with $\mu=\lambda p/r$,
\begin{align}
f_p\bigl(\tfrac{T(x^*)}{p}\bigr)^\lambda f_q\bigl(\tfrac{U(x^*)}{q}\bigr)^{1-\lambda} &=  
\frac{1}{\|f\|_p^{\lambda p}\|f\|_q^{(1-\lambda)q)}} f\bigl(\tfrac{T(x^*)}{p}\bigr)^{\lambda  p}f\bigl(\tfrac{U(x^*)}{q}\bigr)^{(1-\lambda)q}\\
&\leq \frac{1}{\|f\|_p^{\lambda p}\|f\|_q^{(1-\lambda)q)}} f\bigl(\tfrac{\lambda T(x^*)+(1-\lambda)U(x^*)}{r}\bigr)^{r}\\
&=\frac{\|f\|_r^r}{\|f\|_p^{\lambda p}\|f\|_q^{(1-\lambda)q)}}f_r\bigl(\tfrac{\lambda T(x^*)+(1-\lambda)U(x^*)}{r}\bigr).
\end{align}
Using the arithmetic-geometric mean inequality~\eqref{amgm} and integrating the density~\eqref{densitylambda} over $x^*\in\R^n$, one obtains
\begin{equation}
(p^\lambda q^{1-\lambda})^n  \|f\|_p^{\lambda p}\|f\|_q^{(1-\lambda)q)} \leq r^n \|f\|_r^r.
\end{equation} 
Taking the logarithm yields the announced concavity.
\end{proof}

As a side result it is interesting to note that we have obtained a simple transportation proof of the following varentropy bound:
\begin{Corollary}[Varentropy Bound~\cite{FradeliziMadimanWang16}]\label{varentropy}
One has
$\mathrm{Var} \log f(X_p) \leq n/p^2$, that is, $\mathrm{Var} \log f_p(X_p) \leq n$.
\end{Corollary}

\begin{proof}
Since $n\log p + (1-p) h_p(X)$ is concave, one has $\frac{\partial^2}{\partial p^2} \bigl( n\log p + (1-p) h_p(X)\bigr)\leq 0$, that is,
\begin{equation}
\frac{\partial^2}{\partial p^2} \bigl((1-p) h_p(X)\bigr)\leq \frac{n}{p^2}.
\end{equation}
Differentiating twice using Leibniz's product rule and plugging the identity~\eqref{identitydiff}, the l.h.s. of this inequality becomes
\begin{equation}
-\frac{1}{1-p}\frac{\partial D(f_p\|f)}{\partial p} = \frac{\partial}{\partial p} \int f_p \log f
= \int f_p \log^2 f - \bigl(\int f_p \log f\bigr)^2.\qedhere
\end{equation}
\end{proof}

As another easy consequence of Lemma~\ref{logconcaveentropyconcave}, since $n\log p + (1-p) h_p(X)$ is concave and vanishes for $p=1$, the slopes $\frac{n\log p + (1-p) h_p(X)-0}{p-1}$ are nonincreasing in $p$. In other words $h_p(X) + n \frac{\log p }{1-p}$ is nondecreasing. Therefore:
\begin{Corollary}[Marsiglietti and Melbourne~\cite{MarsigliettiMelbourne18}]\label{mmincrease}
If $p<q$ then for any $X$ with log-concave density, $h_p(X)+n \frac{\log p}{1-p}\leq h_q(X)+n \frac{\log q}{1-q}$. 
\end{Corollary}

We can now use Lemma~\ref{charact} and Corollary~\ref{repi1m} to obtain Rényi EPIs for orders $r<1$.
Since all $r_i$ are $>r$, by~Corollary~\ref{mmincrease},
\begin{equation}
h_{r_i}(X)+n \frac{\log r_i}{1-r_i} \geq h_r(X)+n \frac{\log r}{1-r} \qquad (i=1,2,\ldots,m). 
\end{equation}
Plugging this into~\eqref{repi1mineq} one obtains
\begin{align}
h_r\Bigl( \sum_{i=1}^m\sqrt{\lambda_i}X_i\Bigr) -  \sum_{i=1}^m \lambda_i h_{r}(X_i)
&\geq n \bigl(\frac{\log r}{1-r}- \sum_{i=1}^m \lambda_i\frac{\log r_i}{1-r_i}\bigr)
+ \frac{n}{2} r' \Bigl( \frac{\log r}{r}-\sum_{i=1}^m\frac{\log r_i}{r_i}\Bigr)\\
&=\frac{n}{2} r' \Bigl( \sum_{i=1}^m\frac{\log r_i}{r_i}-\frac{\log r}{r}\Bigr) \label{repi2mineqlogc}
\end{align}
where we have used that $\lambda_i=r'/r'_i$ for $i=1,2,\ldots,m$. Notice that the r.h.s. of~\eqref{repi2mineqlogc}
for $r<1$ ($r'<0$) is the opposite of that of~\eqref{repi2mineq} for $r>1$ ($r'>0$). However, since $r'$ is now negative, the r.h.s. is exactly equal to $\frac{n}{2}A(\lambda)$ which is still convex and negative.
For this reason, the proofs of the following theorems for $r<1$ are such repeats of the theorems obtained previously for $r>1$.

\begin{Theorem}
The Rényi EPI~\eqref{repic} for log-concave densities holds for $c=r^{-r'/r}\bigl(1-\frac{1}{mr'}\bigr)^{1-mr'}$ and $r<1$.
\end{Theorem}
\begin{proof}
Identical to that of Theorem~\ref{thmrepic} except for the change $|r'|=-r'$ in the expression of $A(\lambda)$.
\end{proof}

\begin{Theorem}
The Rényi EPI~\eqref{repialpha} for log-concave densities holds for $r<1$ and
$\alpha=\Bigl(1+{|r'| \frac{\log_2 r}{r} +(2|r'|+1) \log_2 \bigl(1+\frac{1}{2|r'|}\bigr)}  \Bigr)^{-1}$
and this value of $\alpha$ cannot be improved using the method of this paper by making it depend on $m$.
\end{Theorem}
\begin{proof}
Identical to that of Theorem~\ref{thmrepialpha} except for the change $|r'|=-r'$ in the expression of $A(\lambda)$.
\end{proof}

\begin{Remark}
The case $m=2$ yields $\alpha=\frac{1-r}{(r+1)\log_2(r+1)-r\log_2 r-2r}$ which was found by Marsiglietti and Melbourne~\cite{MarsigliettiMelbourne18}.  Again the exponent of the Theorem is not improved for $m>2$. 
\end{Remark}

\begin{Theorem}
The Rényi EPI~\eqref{repig} for log-concave densities holds for $c=\Bigl(mr^{-r'/r}\bigl(1-\frac{1}{mr'}\bigr)^{1-mr'}\Bigr)^\alpha / m$ where $r<1$ and $0<\alpha<1$.
\end{Theorem}
\begin{proof}
Identical to that of Theorem~\ref{thmrepig} except for the change $|r'|=-r'$ in the expression of $A(\lambda)$.
\end{proof}

\section{Conclusion}\label{conclusion}

This article provides a comprehensive framework to derive known Rényi entropy power inequalities (with shorter proofs), and prove new ones. The framework is based on a transport argument from normal densities and a change of variable by rotation. Only basic properties of Rényi entropies are used in the proofs.

In particular, the $\alpha$-modification of the EPI is recovered for two or more independent variables for Rényi entropy orders $>1$ as well as for orders $<1$. Also, the Rényi EPI with multiplicative constant~$c$ is extended to Rényi entropy orders $<1$, and a more general formulation with both exponent~$\alpha$ and constant~$c$ is obtained for all orders.  In passing, a simple proof using normal transport of a recent sharp varentropy bound was obtained for log-concave densities.

As a perspective, the methods developed in this paper can perhaps be generalized to obtain reverse Rényi entropy power inequalities (see e.g., the discussion in~\cite{Li18}).

\vspace{6pt} 




%


\reftitle{References}



\end{document}